# Control of nonreciprocal charge transport in topological insulator/superconductor heterostructures with Fermi level tuning and superconducting-layer thickness

Soma Nagahama[1], Yuki Sato[2], Minoru Kawamura[2], Ilya Belopolski[2], Ryutaro Yoshimi[3], Atsushi Tsukazaki[1,4], Naoya Kanazawa[5], Kei S Takahashi[2], Masashi Kawasaki[1,2], and Yoshinori Tokura[1,2,6]

[1]Department of Applied Physics and Quantum-Phase Electronics Center (QPEC), The University of Tokyo, Tokyo 113-8656, Japan

[2]RIKEN Center for Emergent Matter Science (CEMS), Wako 351-0198, Japan

[3]Department of Advanced Materials Science, The University of Tokyo, Kashiwa 277-8561, Japan

[4]Institute for Materials Research (IMR), Tohoku University, Sendai 980-8577, Japan

[5]Institute of Industrial Science, The University of Tokyo, Tokyo 153-8505, Japan and

[6]Tokyo College, The University of Tokyo, Tokyo 113-8656, Japan

**Abstract**

Nonreciprocal charge transport (NCT) is defined as a phenomenon where electrical resistance depends on the current direction. It has been drawing much attention because it sensitively reflects the symmetry breaking of material systems. A topological insulator (TI)/superconductor (SC) heterostructure where the topological surface state (TSS) of the TI layer is proximitized with the SC layer is one such system that presents a sizable NCT due to a large spin-orbit coupling and superconductivity. Here, we report a control of the magnitude and sign of NCT: reversal of the direction of NCT by tuning the Fermi energy of TSS of the TI layer with respect to the charge neutral point by systematic regulation of Sb composition $x$ in a TI/SC heterostructures of $(Bi_{1-x}Sb_x)_2Te_3/FeSe_{0.1}Te_{0.9}$. The result is consistent with the model of a TSS proximitized with superconductivity. Furthermore, we find a significant enhancement of the magnitude of NCT in the TI/SC heterostructures by reducing the thickness of the SC layer. The enhancement can be ascribed to the inversion-symmetry breaking of the $FeSe_{0.1}Te_{0.9}$ SC-layer itself adjacent to the TI layer. Our results highlight the essential role of the TSS for exhibiting NCT and offer new knobs to control the direction and magnitude of NCT.

**Introduction**

Nonreciprocal charge transport (NCT), a phenomenon of the electrical resistance depending on the current direction, has been intensively studied in materials where both the spatial inversion symmetry and time reversal symmetry are simultaneously broken, *i.e.* in noncentrosymmetric materials under magnetic fields or with spontaneous magnetizations [1–3]. So far, NCT has been reported in many noncentrosymmetric materials such as Rashba-type semiconductors [4, 5], transition metal dichalcogenides [6], kagome superconductors [7], polar superconductors [8], and chiral magnets [9]. Designed heterostructures of topological insulator (TI)/superconductor (SC), which are drawing attention as a platform for topological SC [10], are also reported to exhibit NCT [11,12]. Such a superconducting system is expected to show large NCT, because the characteristic energy scale reduces from the Fermi energy $E_F$ for the normal state to the superconducting gap $\Delta$, amplifying perturbative nonlinear effects [6,13]. Superconducting diode effect [14,15], intensively studied recently, can also be regarded as one of the NCT phenomena, where the critical current for the breakdown of the superconductivity depends on the current directions. While the superconducting diode effect requires a large bias current exceeding the critical current of superconductivity, the NCT investigated in Refs. 11 and 12 occur near the critical temperature of superconductivity and are signified by the linear change in the resistance with current.

The microscopic mechanism of NCT in noncentrosymmetric superconductors near the superconducting transition temperature has been studied theoretically by Hoshino *et al*. [13], and the TI/SC heterostructure is specifically discussed as one of the representative systems. They consider a two-dimensional superconductivity in the TSS of three-dimensional TI proximitized with the *s*-wave SC and present the analysis of charge transport above the Berezinskii-Kosterlitz-Thouless (BKT) transition temperature ($T_{\text{BKT}}$) where dynamics of thermally induced vortex and anti-vortex pairs produces voltage. The vortex motion driven by the electrical current becomes nonreciprocal reflecting the inversion symmetry breaking of the TSS under the in-plane magnetic field. The nonreciprocal resistance in the system can be written as

$$R = R_0[1 + \gamma \hat{z} \cdot (\boldsymbol{I}/w \times \boldsymbol{B})] \qquad (1)$$

up to the first order of magnetic field $B$ and current $I$. Here, $R_0$ is the resistance under the zero magnetic field, $\hat{z}$ is the polar axis normal to the TSS, $w$ is the sample width, and $\gamma$ is the nonreciprocal coefficient describing the magnitude of NCT. TSS is usually modeled by a linear dispersion Dirac Hamiltonian. Importantly, however, the linear dispersion alone as in Fig. 1(a) does not produce any nonreciprocal response, and one has to take into account the parabolic term (Fig. 1(b)) or hexagonal warping term (Fig. 1(c)) in the TSS dispersion. The nonreciprocal resistance is described by the nonreciprocal modulation of $T_{\text{BKT}}$ by a magnetic field and a bias current, as $\Delta T_{BKT} \propto IB$. Hence, the coefficient $\gamma$ diverges as $\gamma = \alpha(T - T_{\text{BKT}})^{-3/2}$ as lowering temperature $T$ to approach $T_{\text{BKT}}$ [11-13], provided Halperin-Nelson formula of the resistance $R$ above $T_{\text{BKT}}$, $R = R^* exp\left(-2b\left(\frac{T_0-T}{T-T_{BKT}}\right)^{1/2}\right)$ [16]. Thus, the prefactor $\alpha$ represents the magnitude of nonreciprocity. The important outcome of the model of the TSS proximitized with superconductivity is that the coefficient $\alpha$ depends on the sign of $E_{\text{F}}$ measured from the charge neutral point of TSS in the case where the parabolic term is added to the ordinary linear dispersion, while it does not in the case where the hexagonal warping term is added. In the earlier studies of NCT, little attention has been paid to the control of the direction of the reciprocity or the sign of $\gamma$. Since $E_F$ of the TSS can be tuned experimentally, the TI/SC system can provide a unique opportunity to investigate the direction of NCT.

In this paper, we report NCT near the superconducting transition temperature in thin films of TI/SC heterostructure $(Bi_{1-x}Sb_x)_2Te_3/FeSe_{0.1}Te_{0.9}$ grown by molecular beam epitaxy (MBE). We observed that the sign of the NCT switches when $E_F$ crosses the charge neutral point by tuning the Sb composition $x$. The sign reversal is consistent with the theoretical prediction assuming a parabolic dispersion term in the dispersion relation of the TSS. Furthermore, we found that the magnitude of NCT is largely enhanced by thinning the superconducting layer, suggesting the dominant contribution from the noncentrosymmetric SC layer itself. Our findings thus not only contribute to understanding the microscopic mechanism of NCT in the designed TI/SC heterostructure but also offer a new knob to control both the direction and amplitude of NCT.

**Synthesis of superconductor/topological insulator heterostructures**

To study the $E_F$ dependence of the nonreciprocal response experimentally, we prepared a series of heterostructure films of $(Bi_{1-x}Sb_x)_2Te_3/FeSe_{0.1}Te_{0.9}$ as schematically depicted in Figs. 1(d) and 1(e). $(Bi_{1-x}Sb_x)_2Te_3$ is a well-established three-dimensional topological insulator with a single Dirac cone dispersion on each topological surface state (TSS). In the $(Bi_{1-x}Sb_x)_2Te_3$ (BST) layer, it is possible to control $E_F$ by adjusting the Sb composition $x$ [17]. For the superconducting layer, we adopted $FeSe_{0.1}Te_{0.9}$ (FST), which belongs to one of the Fe-based superconductor families with the simplest crystal structure (Fig. 1(d)) and has a relatively high superconducting transition temperature $T_c$ of around 12 K [18]. Both BST and FST have van der Waals structure, allowing us to grow BST with $0 \leq x \leq 1$ on top of FST using MBE. We used CdTe (100) as a substrate and fixed the composition of FST, *i.e.*, $FeSe_{0.1}Te_{0.9}$, where the optimal $T_c$ is obtained [19]. In the heterostructure films, the thickness of each layer was fixed at 10 nm and 8 nm for BST and FST, respectively (Fig. 1(e)), unless otherwise stated. We determined these film thicknesses so that they are thin enough to obtain a large current density at the interface but thick enough to assure the presence of the TSS in the BST layer [20]. All the samples were passivated by depositing a capping layer of $AlO_x$ (3 nm) using atomic layer deposition. Figure 1(f) shows the x-ray diffraction characterization result where no peak other than the target materials of BT and FST and substrate were observed, indicating no crystalline impurity phase discerned in the film. Atomic force microscopy (AFM) images show reasonably flat surface, supporting the cleanness of the films (Supplementary Fig. S1) [21]. To measure NCT, we fabricated the rectangular-shaped micro-scale devices and attached the Au/Ti electrodes (Fig. 1(g)). Both the width of the samples and the distance of the voltage terminals are 100 μm.

**Nonreciprocal charge transport measurement**

To confirm the two-dimensional nature of superconductivity in our system, we first conduct charge transport measurements on FST single-layer (8 nm) at low temperatures down to 2 K. Figure 1(h) shows a logarithmic-scale plot of current ($I$)-voltage ($V$) curves of the FST single-layer. The voltage $V$ follows the power-law dependence on $I$ at low temperatures, which is a characteristic of two-dimensional superconductivity. Theoretically, the power $p$ ($V \sim I^p$) reaches 3 at $T_{BKT}$, which is evaluated to be 5.6 K in this sample. As shown in Fig. 1(i), the temperature dependence of resistance is well fitted by Halperin-Nelson formula [16] with a similar value of $T_{BKT} = 5.5$ K, which also signifies the two-dimensional superconductivity in the FST. Next, we turn our attention to nonreciprocal resistance. To obtain $\gamma$, we apply an AC current $I = \sqrt{2}I_0 \sin \omega t$ in the $x$ direction and a magnetic field $B$ in the $y$ direction, measure the voltage drop at a frequency of $2\omega$, and obtain the second harmonic resistance $R_{2\omega} = \text{Im}[V_{2\omega}/I_0]$. We also obtain primary resistance $R_0$ measured at a small current ($I = 10$ μA). Then, $\gamma$ is yielded from the slope of $R_{2\omega}/R_0$ versus $B$ curve as $\gamma = -\sqrt{2}R_{2\omega}w/R_0BI_0$. We confirmed that $R_{2\omega}$ is linearly dependent on the current and independent of the frequency (Supplementary Fig. S2) [21]. Hereafter, we use $I_0 = 0.2$ mA and $\omega/2\pi = 13$Hz in this paper. Figure 1(j) shows the magnetic field dependence of $R_{2\omega}/R_0$ for the BT/FST heterostructure near the BKT transition ($T = 9.6$ K) and in the normal state ($T = 13.0$ K). The signal is non-zero at $T = 9.6$ K while it is almost zero at $T = 13.0$ K. Near the BKT transition ($T = 9.6$ K), $R_{2\omega}/R_0$ increases linearly with $B$ near the zero magnetic field and is almost antisymmetric with $B$ as anticipated from Eq. (1). Hereafter, we focus on the antisymmetric part of $R_{2\omega}/R_0$ and deal with the anti-symmetrized data against the magnetic field, $R_{2\omega}/R_0 = (R_{2\omega}/R_0(B)-$ $R_{2\omega}/R_0(-B))/2$. While we adopted the AC measurement technique, we note that we could observe the quantitatively comparable nonreciprocal current-voltage characteristic curves by DC measurements (Supplementary Fig. S3 and Supplementary Discussion 1) [21].

**Sign reversal of nonreciprocal resistance**

Figure 2(a) shows the temperature dependence of resistance $R_{1\omega}$ of three representative samples, $Bi_2Te_3$/FST (BT/FST), $Sb_2Te_3$/FST (ST/FST), and FST (8 nm) single-layer film under zero magnetic field. All the three samples undergo superconducting transitions at around 10 K. The zero-resistance temperature in our single-layer FST films is lower than that reported in the previous study [19]. It is suggested that the difference is associated with the film thickness and the two-dimensional character of the superconductivity. Also, variation in their zero-resistance temperatures might be attributed to the charge transfer between the FST and BT (ST) layers [22]. Figures 2(b)-(d) show the ratio $R_{2\omega}/R_0$ for the samples. In the former two samples, sizable $R_{2\omega}/R_0$ values are observed under in-plane magnetic fields, while $R_{2\omega}/R_0$ is small in the FST single-layer film. In the BT/FST sample, $R_{2\omega}/R_0$ peaks at around $B = 1$ T and decreases gradually with increasing $B$. In the ST/FST sample, on the other hand, $R_{2\omega}/R_0$ decreases monotonically with $B$. The nonlinearity in the $B$ dependence of $R_{2\omega}/R_0$ may be attributed to a contribution from the higher-order nonlinear phenomena beyond our considerations, such as the field-induced in-plane superconducting vortices [23]. To avoid complications arising from the nonlinearity of $B$ dependence, we hereafter focus on the $B$-linear regime near zero magnetic field. We evaluated the nonreciprocal coefficient $\gamma = -\sqrt{2}R_{2\omega}w/R_0BI_0$ by performing $B$-linear fittings to the data within the low magnetic field region (-0.5 T $\leq B \leq$ 0.5 T). The temperature dependence of $\gamma$ is shown in Figs. 2(e)-(g) for the three samples together with the temperature dependence of $R_0$. The absolute value of $\gamma$ in the FST single-layer film is small compared to those in the other two samples.

The present result indicates that the TSS of the BST layer plays an essential role in the observed NCT

in the TI/SC heterostructures. As shown in Figs. 2(e)-(f), the absolute magnitude of $\gamma$ for BT/FST and

ST/FST increases divergently as the temperature approaches $T_{BKT}$, accompanied by vanishing $R_0$. The temperature dependence is well fitted (black curves in Figs. 2(e) and 2(f)) by the formula $\gamma = \alpha(T -$

$T_{BKT})^{-3/2}$, with fitting parameters $\alpha$ and $T_{BKT}$, in accord with the theoretical prediction [13].

The most remarkable difference between BT/FST and ST/FST lies in the sign of $\gamma$; BT/FST exhibits the negative $\gamma$, while ST/FST shows the positive $\gamma$. The sign reversal indicates that the direction of

NCT is opposite between BT/FST and ST/FST. To understand the origin of the sign reversal, we systematically investigated a series of $(Bi_{1-x}Sb_x)_2Te_3$/FST films with varying $x$ between 0 (BT) and 1

(ST) (see Supplementary Fig. S4) [21]. Figure 2(h) shows the change of the coefficient $\alpha$ as a function of the composition $x$. Here, we use $\alpha$ instead of $\gamma$, because $\alpha$ is more suitable for comparing different samples with different $T_{BKT}$. The absolute value of $\alpha$ in $x = 0$ is remarkably larger than those in other compositions. We speculate that the possible reason is the clean BT/FST interface due to the good lattice matching between $\sqrt{3}a_{BT}$ and $2a_{FST}$ [24]. Importantly, the sign reversal of $\alpha$ occurs between $x$

= 0.85 and 0.95. This composition is close to the point where the sign of the inverse Hall coefficient

($1/R_H$) also flips in BST films grown on InP(111) substrates (Fig. 2(i)). Thus, the sign reversal of $\alpha$ in BST/FST appears to occur near the charge neutral point (CNP) of the TSS. The small deviation in $x$ between the critical compositions for the sign reversal of $\alpha$ and the CNP of the TSS may be attributed to an electron transfer from FST to BST (Supplementary Figs. S5 and S6, and Supplementary Discussion 2) [21]. We note that estimation of carrier density in the BST/FST heterostructure is difficult because of the multi-band character of the Fe(Se,Te) and the combination of two different materials, though it is possible for BST single-layer film (Supplementary Fig. S7) [21]. According to the theory of the TSS proximitized with superconductivity [13], the sign reversal can occur when the TSS has a quadratic dispersion in addition to the linear dispersion. It is to be noted that in the $k \rightarrow 0$ limit around the CNP, the parabolic term ($k^2$ term) should be dominating over the hexagonal warping term ($k^3$ term). Therefore, our observation of the sign reversal of $\alpha$ around the CNP confirms the relevance of the TSS to the NCT and supports the essential contribution from the parabolic term in the TSS.

**Magnetic field angle dependence**

To elaborate on the intrinsic origin of the sign reversal in the observed $E_F$-dependent NCT, we investigate the magnetic field angle dependence of the nonreciprocal resistance. For this purpose, we used the samples of BT/FST and ST/FST. We define the magnitude of NCT in general magnetic field angle as $\gamma'(\theta, \phi) = \sqrt{2}R_{2\omega}w/R_0BI_0$, where $\phi$ is the azimuthal angle and $\theta$ is the polar angle. Here, note that $\gamma'(\pi/2, \pi/2) = \gamma$. First, we fixed the magnetic field direction at an azimuthal angle $\phi$ in the $xy$ plane and measured the magnetic field dependence of $R_{2\omega}/R_0$ (Figs. 3(a) and 3(b)). Then, $\gamma'$ is evaluated from the slope of $R_{2\omega}/R_0$-$B$ curves. By repeating this procedure for various angles $\phi$ and plotting $\gamma'$ against $\phi$, we obtain Fig. 3(c). The temperature was fixed to $T = 9.5$ K and 11.8 K for BT/FST and ST/FST, respectively. The data for both BT/FST and ST/FST follow the $\sin\phi$ dependence, as expected from Eq. (1). The sign of nonreciprocal signal is always negative in BT/FST and positive in ST/FST at any $\phi$. Next, we conducted similar measurements under the magnetic field rotation in the $yz$ plane with the polar angle $\theta$ (Figs. 3(d)-(f)). No sign reversal was observed either in the $\theta$-scan measurement. The magnetic field angle dependence clearly confirms that the sign reversal accompanied by the composition tuning cannot be attributed to the misalignment of the magnetic field. Nevertheless, the $\theta$ dependence data do not follow the simple $\sin\theta$ curves for both for BT/FST and ST/FST. We speculate that the deviation from the simple $\sin\theta$ dependence may be related to the imbalance of the number of vortices and anti-vortices in the presence of the out-of-plane magnetic field.

**Film thickness dependence**

Next, we examine the FST layer thickness dependence of the NCT. In the BST/FST heterostructures, the electrical current is distributed into BST, FST, and their interface. Naively thinking, reducing the FST layer thickness results in the increase of the proportion of the interface current, leading to more pronounced NCT. To study the thickness dependence, we prepared a series of samples of $Bi_2Te_3$(10

nm)/ FST($t$ nm) and Sb$_2$Te$_3$(10 nm)/ FST($t$ nm) heterostructures with thickness $t$ = 2, 4, and 8 nm. All the films show superconducting transitions, and $\gamma$ diverges following the $\gamma = \alpha(T - T_{BKT})^{-3/2}$ dependence, as shown in Fig. 4(a). In the case of BT/FST, the absolute magnitude of $\alpha$ increases about seven times as $t$ decreases from 8 nm to 2 nm (Fig. 4(b)). In the case of ST/FST, however, the sign of $\alpha$ changes from positive to negative, and the absolute magnitude of $\alpha$ increases by an order of magnitude by reducing $t$ from 8 nm to 2 nm (Fig. 4(b)). The value of $\alpha = 1.4\times10^{-2}$ T$^{-1}$A$^{-1}$mK$^{1.5}$ for BT/FST (2nm) exceeds the previously reported value in Bi$_2$Te$_3$/FeTe [12]. Although the superconducting layer in Bi$_2$Te$_3$/FeTe is uncertain and direct comparison is a subtle issue, we conclude that the thickness of the superconducting layer is one of the important factors to enhance NCT. The sign reversal in ST/FST by thinning the FST layer cannot be explained by the simple model of the TSS proximitized with superconductivity, suggesting the different mechanism at work. One possible scenario is Rashba-type spin splitting of FST. In the bulk form, FST does not have a polar structure and Rashba-type spin splitting is not accepted. However, in our heterostructures, FST is sandwiched between BST and CdTe(100) substrate and is subject to the inversion symmetry breaking. The broken inversion symmetry can make the FST layer a Rashba-type superconductor where the NCT is anticipated under the in-plane magnetic field [25]. Indeed, in the case of a Bi$_2$Te$_3$/FeTe heterostrucure, electrons are transferred from FeTe layer to Bi$_2$Te$_3$ layer through the interface because of the smaller work function in FeTe than in Bi$_2$Te$_3$ [26,27] and electron transfer from FST to Bi$_2$Te$_3$ (Sb$_2$Te$_3$) is also expected(Supplementary Figs. S5 and S6 and Supplementary Discussion 2) [21,28-30]. In TI/SC heterostructure, therefore, FST layer whose thickness is comparable to the electron screening length (0.5 nm in bulk FeSe$_{0.55}$Te$_{0.45}$ [31]) is under the influence of strong potential gradient. Further careful investigation including the electronic structure of the whole heterostructure would be necessary to elucidate the mechanism of the nonreciprocity in these very thin films viewed as the polar superconductor or the possible topological superconductor.

**Conclusion**

We fabricated the BST/FST heterostructure and observed the sign reversal of NCT around the charge neutral point in BST which is attributed to the contribution from the parabolic term in TSS. Thus, the TSS proximitized with superconductivity contributes dominantly to NCT when FST layer is as thick as 8 nm. We also confirmed this sign reversal is intrinsic to the TI/SC system by rotating magnetic field angle. Finally, by making the constituent FST layer thinner, we observed the further enhancement of the magnitude of NCT possibly from FST layer itself whose centrosymmetry is broken. Besides, the superconducting NCT signals of this origin can be much larger than those from the TSS at the interface. We have demonstrated a control over the direction and magnitude of NCT. Our findings may pave the way to developing novel superconducting diode devices arising from an admixed *p*-wave character of the superconductivity.

**Acknowledgement**

This work was supported by JSPS KAKENHI Grants (No.22H04958, No. 24K17020, No. 22K18965, No. 23H04017, No. 23H05431, No. 23H05462, No. 23H04862, 24H00417, and 24H01652), JST FOREST (Grant No. JPMJFR2038), JST CREST (Grant No. JPMJCR1874 and No. JPMJCR23O3), Mitsubishi Foundation, Sumitomo Foundation, the special fund of Institute of Industrial Science, The University of Tokyo, and the RIKEN TRIP initiative (Many-body Electron Systems). This research was conducted in part using the AtomWork provided by the Materials Data Platform (MDPF) of the National Institute for Materials Science (NIMS).

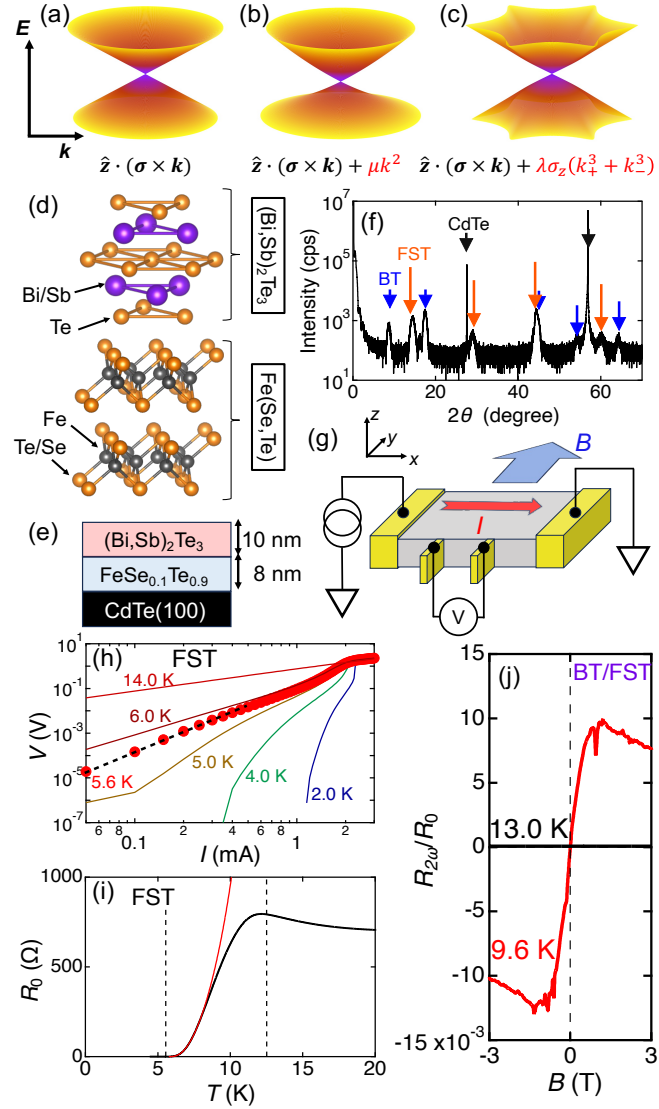

FIG. 1. (a)-(c) Dispersion of topological surface state (TSS) (a) without higher order terms, (b) with a parabolic term, and (c) with a hexagonal warping term. Hamiltonians in each case are shown below the schematics. $k_{\pm} = k_x \pm ik_y$. (d) Crystal structure of Fe(Se,Te) (FST) and $(Bi_{1-x}Sb_x)_2Te_3$ (BST). (e) A schematic of BST/FST heterostructure film grown on a CdTe (100) substrate. (f) X-ray diffraction $2\theta$-$\omega$ scan of a BT/FST heterostructure. Blue, orange, and black arrows indicate the peaks from BT, FST, and CdTe substrate, respectively. (g) A schematic of the AC measurement setup to measure NCT. (h) Log-log plot of $I$-$V$ characteristic curves in an FST single-layer film measured at temperatures of 2.0,

4.0, 5.0, 5.6, 6.0 and 14.0 K. Dashed line is a fit to the $T$ = 5.6 K data, which yields the power $p \sim 3$.

(i) Temperature dependence of resistance in FST single-layer film (black). Red curve shows the fitting

result with the Halperin-Nelson formula, $R = R^* exp\left(-2b\left(\frac{T_0-T}{T-T_{BKT}}\right)^{1/2}\right)$. The yielded fitting parameters

$T_{BKT}$ = 5.5 K and $T_0$ = 12.5 K are shown by dashed lines. (j) Magnetic field dependence of $R_{2\omega}/R_0$ in

BT/FST at 9.6 K in the BKT transition regime (red) and at 13.0 K in the normal state regime (black).

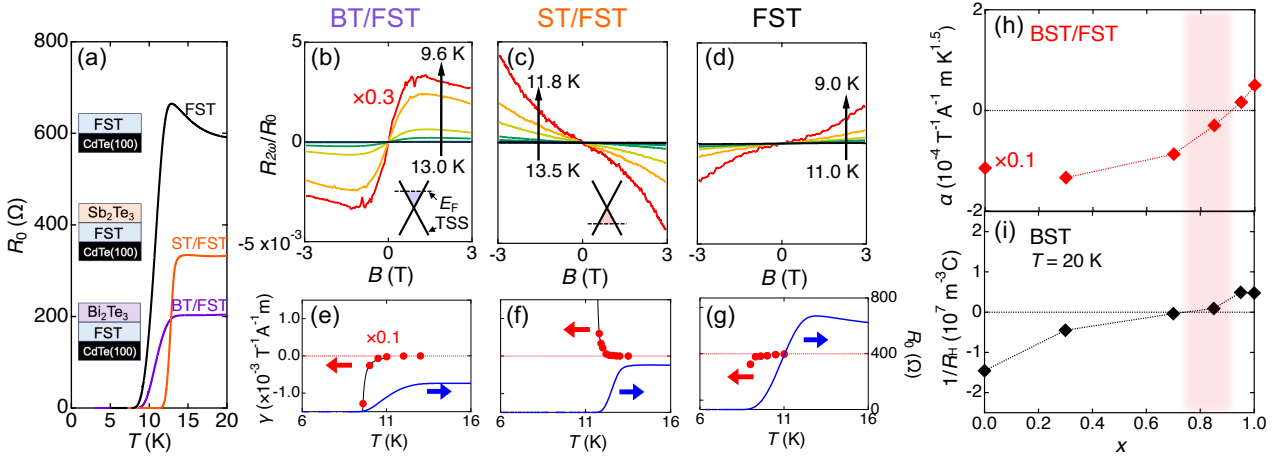

FIG. 2. (a) Temperature dependence of resistance $R_0$ in FST (black), BT/FST (purple), and ST/FST (orange), respectively. Thickness of each film is 8 nm for FST ($FeSe_{0.1}Te_{0.9}$) and 10 nm for BT ($Bi_2Te_3$) and ST ($Sb_2Te_3$). (b)-(d) Magnetic field dependence of $R_{2\omega}/R_{1\omega}$ for (b) BT/FST ($T$ = 9.6, 10.0, 10.5, 11.0, 12.0, and 13.0 K), (c) ST/FST ($T$ = 11.8, 11.9, 12.0, 12.2, 12.4, 12.6, 12.8, 13.0, and 13.5 K), and (d) FST ($T$ = 9.0, 9.3, 9.6, 10.0, 10.5, and 11.0 K). Insets in (b) and (c) show schematics of the band dispersion of topological surface states (TSS) and the corresponding Fermi level ($E_F$) position. (e)-(g) Temperature dependence of nonreciprocal coefficient $\gamma$ (red) and resistance $R_0$ (blue) in (e) BT/FST, (f) ST/FST, and (g) FST. Fitting curves of $\gamma = \alpha(T-T_{BKT})^{-3/2}$ are indicated by black lines. $\gamma$ values in (e) are multiplied by 0.1 for visibility. (h) $x$ dependence of the magnitude of NCT $\alpha$ in the BST/FST heterostructures. The value for $x = 0$ (BT) is multiplied by 0.1 for visibility. (i) $x$ dependence of the inverse Hall coefficient $1/R_H$ for BST (($Bi_{1-x}Sb_x)_2Te_3$, 10 nm) films grown on InP(111) substrates. Red shaded region in (h) and (i) indicates the composition where the sign of $\alpha$ and $1/R_H$ changes.

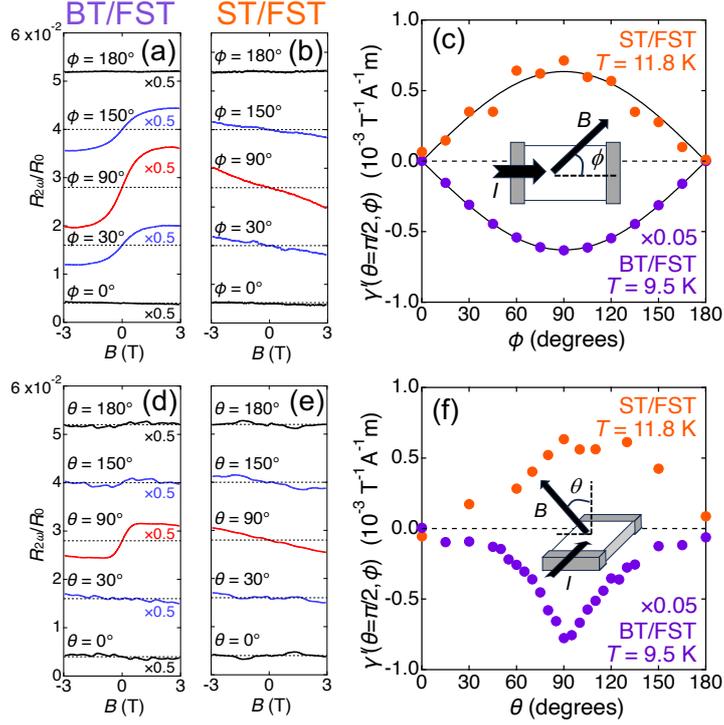

FIG. 3. (a)-(b) Magnetic field dependence of nonreciprocal signal $R_{2\omega}/R_0$ with various azimuthal angle $\phi$ for (a) BT/FST and (b) ST/FST. The zero levels (horizontal dotted lines) for each $\phi$ are arbitral offsets. Magnetic field is rotated within the $xy$ plane with $\phi$ as shown in the inset of (c). The experiment is conducted at the temperature $T$ = 9.5 K for BT/FST and $T$ = 11.8 K for ST/FST. (c) $\phi$ dependence of $\gamma'(\theta, \phi)$. The data for BT/FST and ST/FST are shown in purple and orange, respectively. The data for BT/FST is multiplied by 0.05 for visibility. The solid lines are the fitting curves of $\gamma'(\theta, \phi) = \gamma \sin\phi$. (d)-(e) Magnetic field dependence of nonreciprocal signal $R_{2\omega}/R_0$ with various polar angle $\theta$ in (d) BT/FST and (e) ST/FST. Magnetic field is rotated within the $yz$ plane with $\theta$ as shown in the inset of (f). (f) $\theta$ dependence of $\gamma'(\theta, \phi)$. The zero levels (horizontal dotted lines) for each $\theta$ are arbitral offsets. The data for BT/FST is multiplied by 0.05 for visibility.

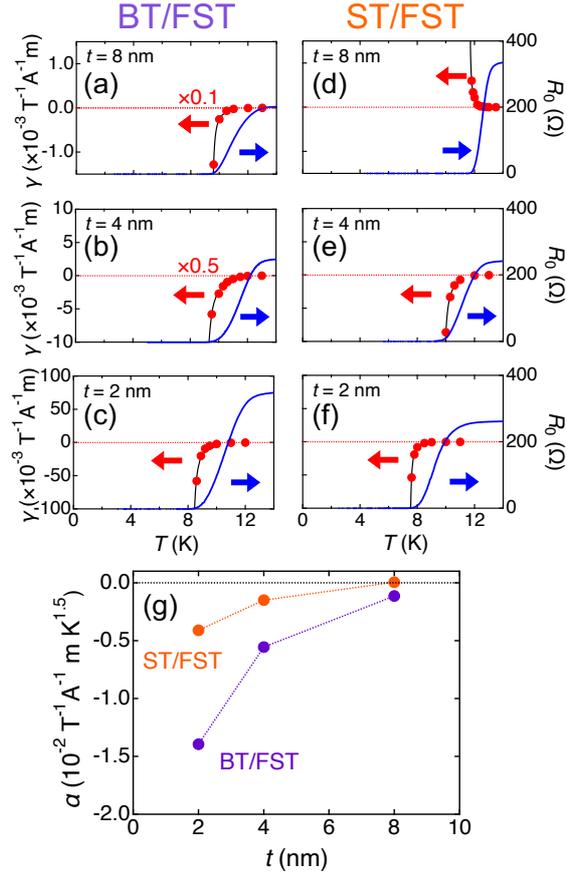

FIG. 4. (a)-(f) Temperature dependence of $\gamma$ and $R_0$. Left columns (a)-(c) are the data for BT/FST, and right columns (d)-(f) for FT/FST. The FST layer thicknesses are $t$ = 2, 4, and 8 nm. $\gamma$ and $R_0$ are shown in red circles and blue lines, respectively. Fitting curves $\gamma = \alpha(T - T_{\text{BKT}})^{-3/2}$ with fitting parameters $\alpha$ and $T_{\text{BKT}}$ are indicated as black lines. (g) FST thickness ($t$) dependence of the magnitude of NCT $\alpha$ for BT/FST (purple) and ST/FST (orange).

**Supplemental Material:**

**Control of nonreciprocal charge transport in topological insulator/superconductor heterostructures with Fermi level tuning and superconducting-layer thickness**

Soma Nagahama[1], Yuki Sato[2], Minoru Kawamura[2], Ilya Belopolski[2], Ryutaro Yoshimi[3], Atsushi Tsukazaki[1,4], Naoya Kanazawa[5], Kei S. Takahashi[2], Masashi Kawasaki[1,2], and Yoshinori Tokura[1,2,6]

[1]Department of Applied Physics and Quantum-Phase Electronics Center (QPEC), The University of Tokyo, Tokyo 113-8656, Japan

[2]RIKEN Center for Emergent Matter Science (CEMS), Wako 351-0198, Japan

[3]Department of Advanced Materials Science, The University of Tokyo, Kashiwa 277-8561, Japan

[4]Institute for Materials Research (IMR),Tohoku University, Sendai 980-8577, Japan

[5]Institute of Industrial Science, The University of Tokyo, Tokyo 153-8505, Japan

[6]Tokyo College, The University of Tokyo, Tokyo 113-8656, Japan

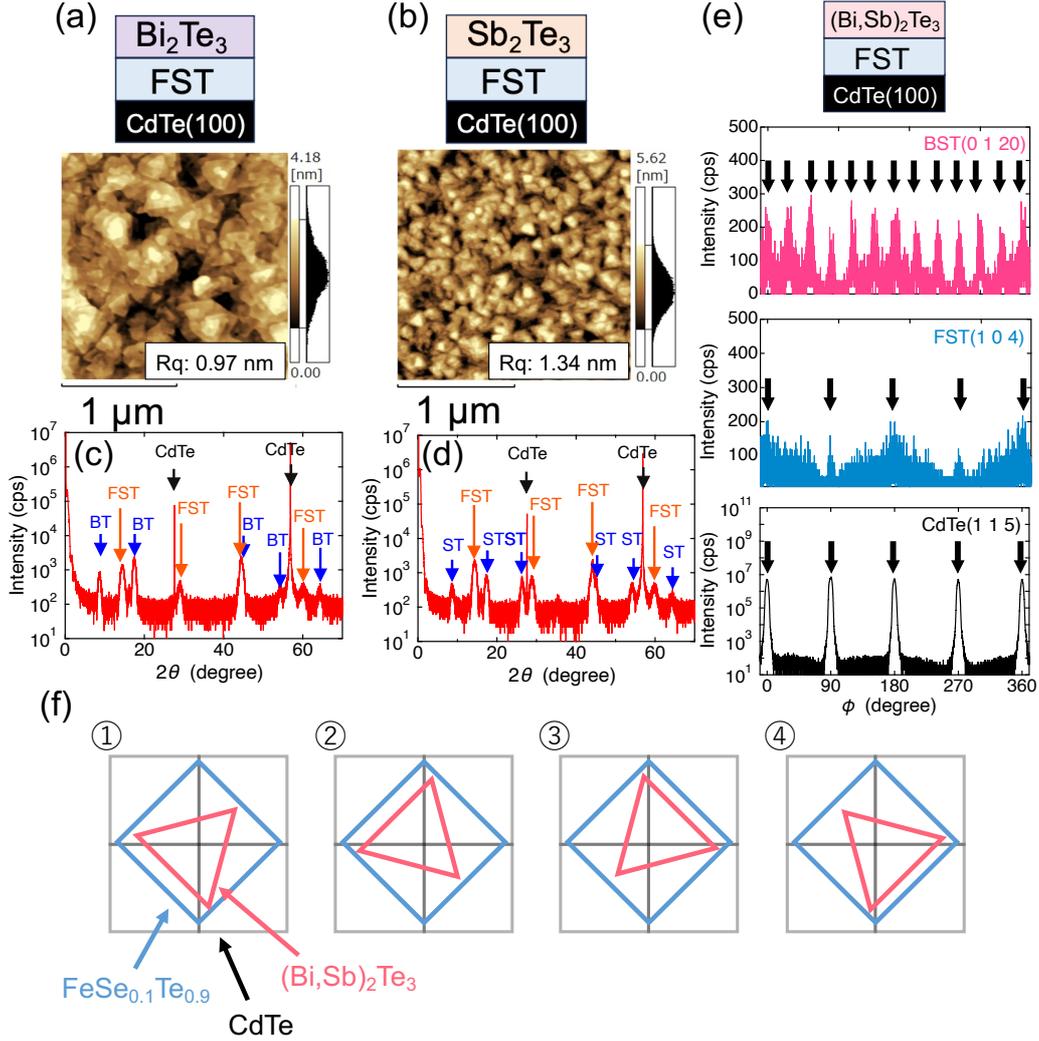

**FIG. S1. Characteristics of (Bi,Sb)₂Te₃/FST heterostructures.** (a)-(b) Atomic force microscope images on the heterostructure thin films (a) BT/FST and (b) ST/FST. The scan area is 2 μm × 2 μm. The Rq value indicates a standard deviation of the sample thickness. (c)-(d) X-ray diffraction $2\theta$-$\omega$ scan data on (c) BT/FST and (d) ST/FST. Blue, orange, and black arrows indicate the peaks from BT(ST), FST, and CdTe substrate, respectively. (e) X-ray diffraction $\phi$ scan data on (Bi,Sb)₂Te₃/FST heterostructure. The Bragg reflection peaks corresponding to (Bi,Sb)₂Te₃(0 1 20), FST(1 0 4), and CdTe(1 1 5) are shown in the top, middle, and bottom panels, respectively. The coincidence of the peak angle positions between FST(1 0 4) and CdTe(1 1 5) indicates that $a$ axis of FST is 45° rotated from that of CdTe, as illustrated in (f). Twelve-fold peaks for (Bi,Sb)₂Te₃ (0 1 20) indicate that (Bi,Sb)₂Te₃ stacks on FST forming 4 different domains, as shown in (f). (f) Possible 4 stacking domain patterns of the heterostructure film. Pink, blue, and black colors illustrate the lattice of (Bi,Sb)₂Te₃, FST, and CdTe, respectively.

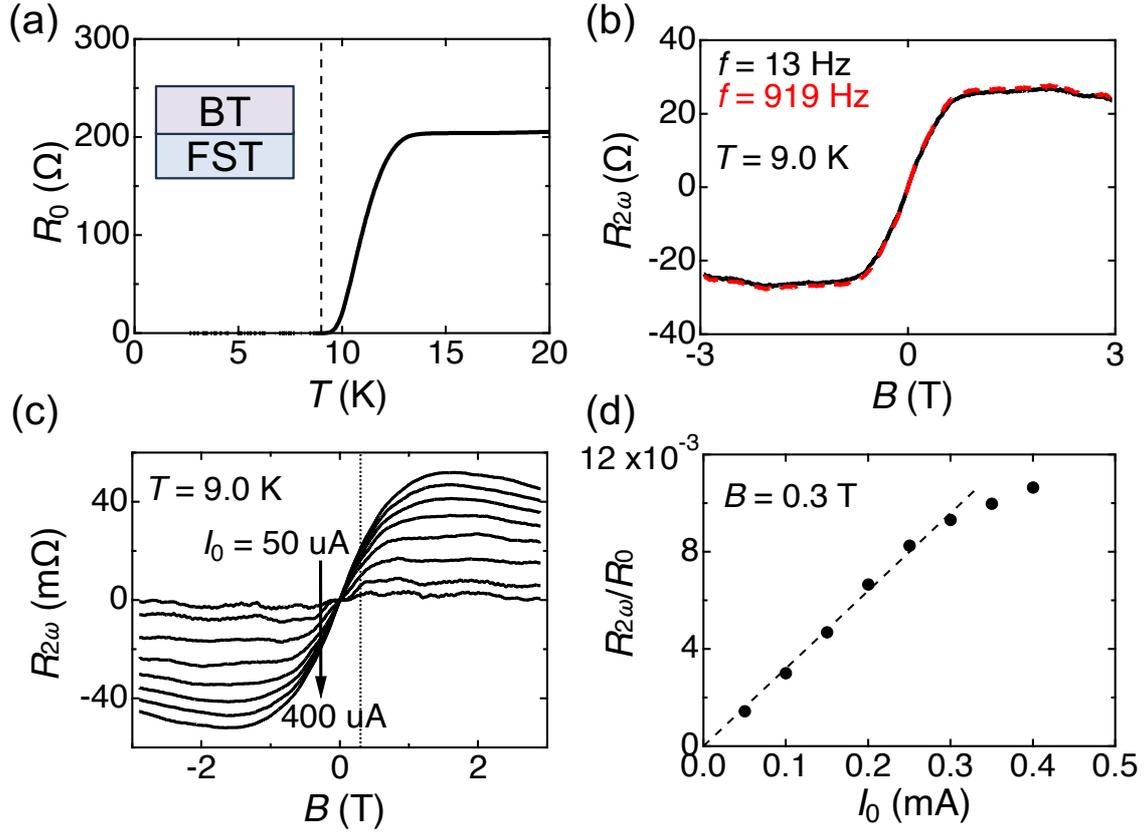

**FIG. S2. Current dependence of nonreciprocal signal.** (a) Temperature dependence of resistance $R_0$ for BT/FST. The dashed line indicates $T = 9.0$ K. (b) Nonreciprocal signal $R_{2\omega}$ measured at $f = 13$ (black solid line) and 919 Hz (red broken line). The temperature is $T = 9.0$ K. (c) Magnetic field dependence of $R_{2\omega}$ at $T = 9.0$ K measured using various currents between $I_0 = 50$ μA and 400 μA with an increment of 50 μA. The vertical dotted line indicates $B = 0.3$ T. (d) Current dependence of $R_{2\omega}/R_0$ at $B = 0.3$ T. With increasing current, $R_{2\omega}/R_0$ increases linearly and turns to saturate above $I_0 \sim 300$ μA. In the main text, we conducted the experiments at $I_0 = 200$ μA.

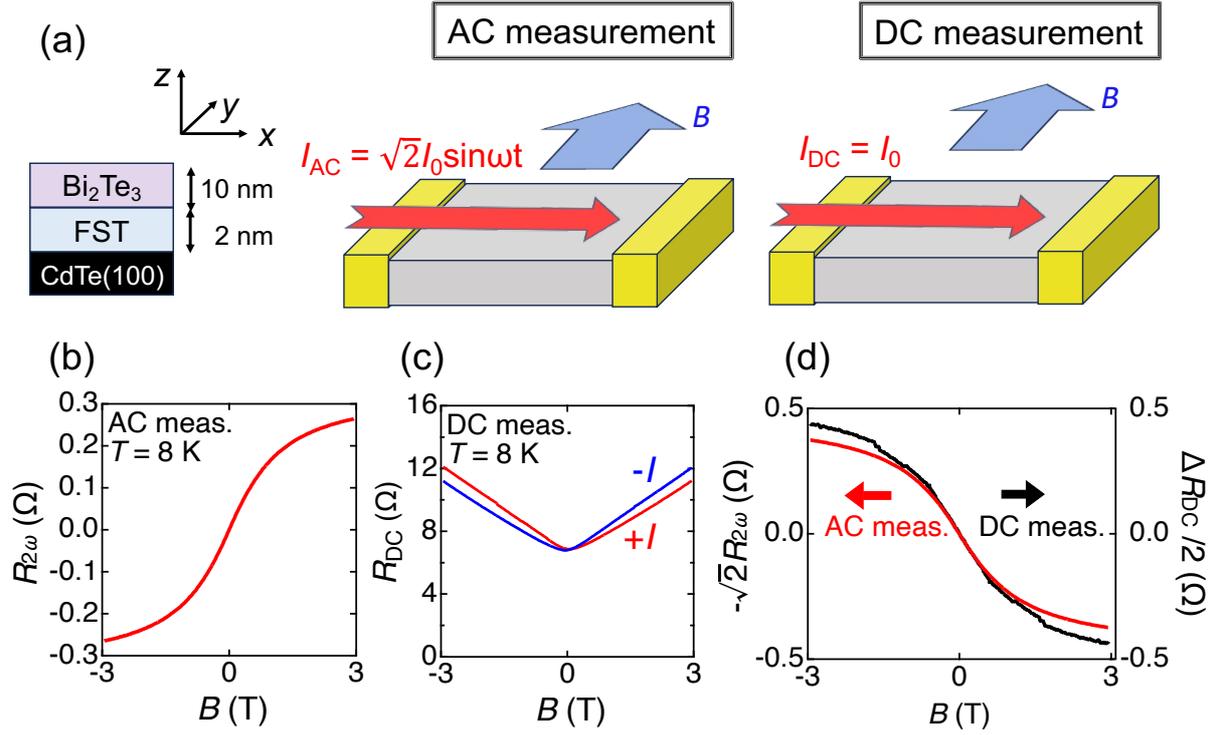

**FIG. S3. Nonreciprocal resistance measured by AC and DC currents.** (a) Schematics of BT/FST film (left), AC measurement (middle), and DC measurement (right). $I_0$ is set to be 200 μA. (b) Magnetic field dependence of $R_{2\omega}$ at $T = 8$ K and $I_0 = 200$ μA. (c) Magnetic field dependence of DC resistance $R_{DC}$ when the current direction is positive (red: $R^+$) and negative (blue: $R^-$). (d) Magnetic field dependence of $-\sqrt{2}R_{2\omega}$ (red) and $\Delta R_{DC}/2$ (black), which are supposed to be equal according to Eq. (1) in the main text. $\Delta R_{DC}$ is defined as $\Delta R_{DC}(B) = R^+(B) - R^-(B)$.

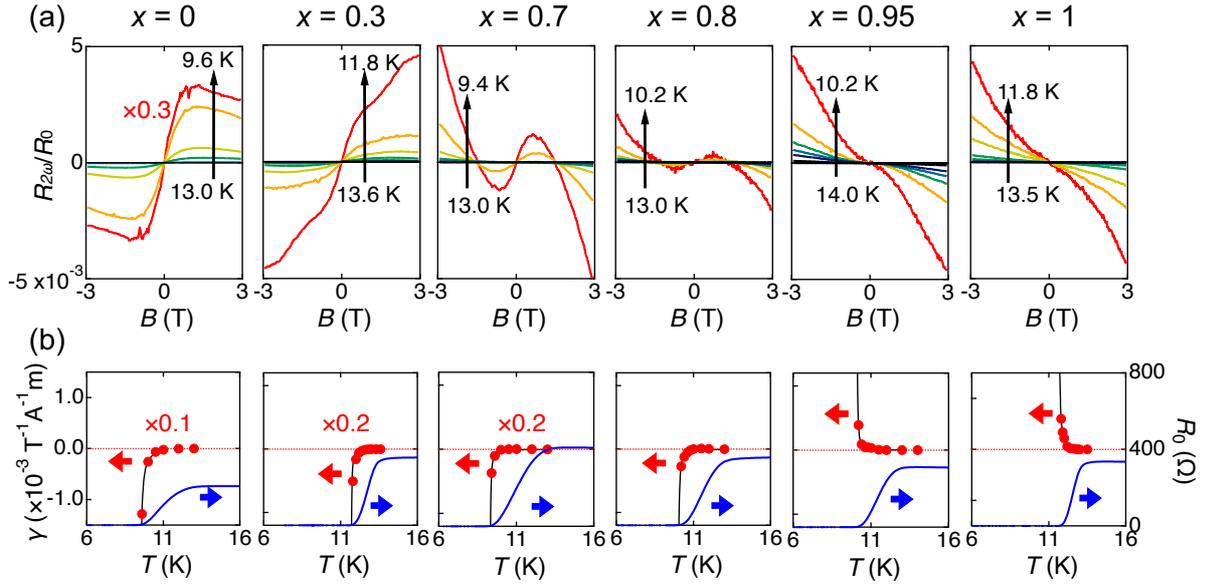

**FIG. S4. Composition dependence of nonreciprocal signal.** (a) Magnetic field dependence of the nonreciprocal signal $R_{2\omega}/R_0$ for various Sb compositions, $x$. (b) Temperature dependence of $\gamma$ (red circles) and resistance $R_0$ (blue lines) for various $x$. Fitting curves of $\gamma = \alpha(T-T_{\mathrm{BKT}})^{-3/2}$ with fitting parameters $\alpha$ and $T_{\mathrm{BKT}}$ are indicated by black lines.

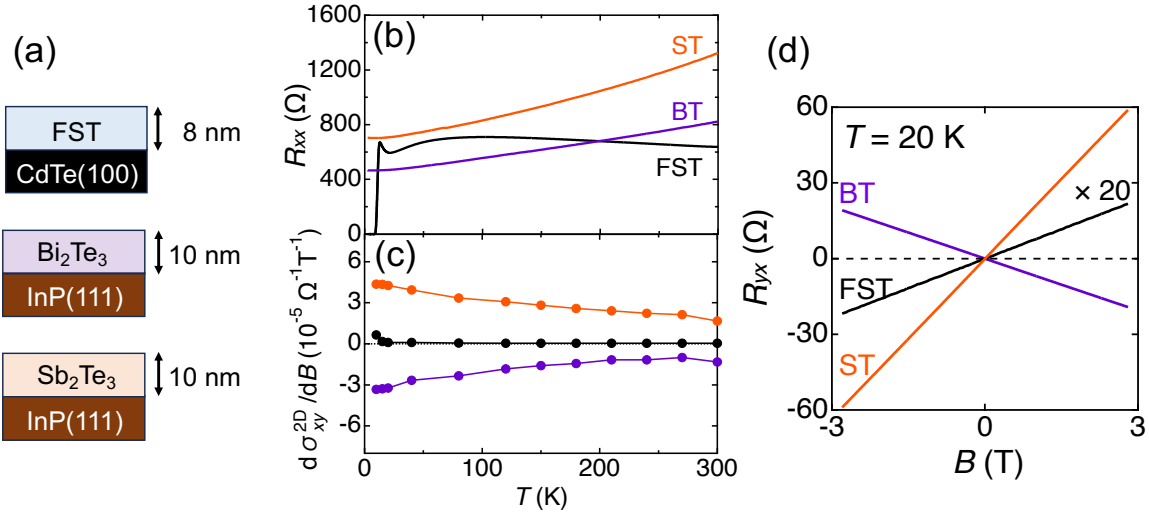

**FIG. S5. Transport properties of single-layer films.** (a) Schematics of the single-layer films. (b)-(c) Temperature dependence of (b) longitudinal resistance $R_{xx}$ and (c) magnetic field derivative of Hall conductance, $d\sigma^{2D}_{xy}/dB$. The derivative is taken at $B = 0$ T. (d) Magnetic field dependence of Hall resistance $R_{yx}$ at $T = 20$ K. The data for FST is multiplied by 20 for visibility.

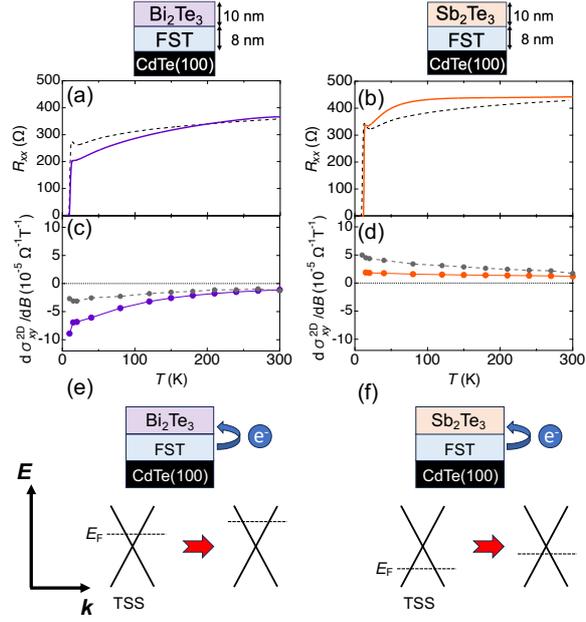

**FIG. S6. Transport properties of heterostructure films.** (a)-(b) Temperature dependence of resistance $R_{xx}$ for (a) BT/FST and (b) ST/FST. (c)-(d) Temperature dependence of $d\sigma^{2D}_{xy}/dB$ for (c) BT/FST and (d) ST/FST. Dotted curves in (a)-(d) are calculated using the data for single-layer BT, ST, and FST films in Fig. S5, assuming parallel circuits. (e)-(f) Schematics of charge transfer and resultant changes in the Fermi energy $E_F$. Due to the electron transfer from FST to BST, $E_F$ shifts away from the charge neutrality point (CNP) in BT/FST, while it comes close to the CNP in ST/FST.

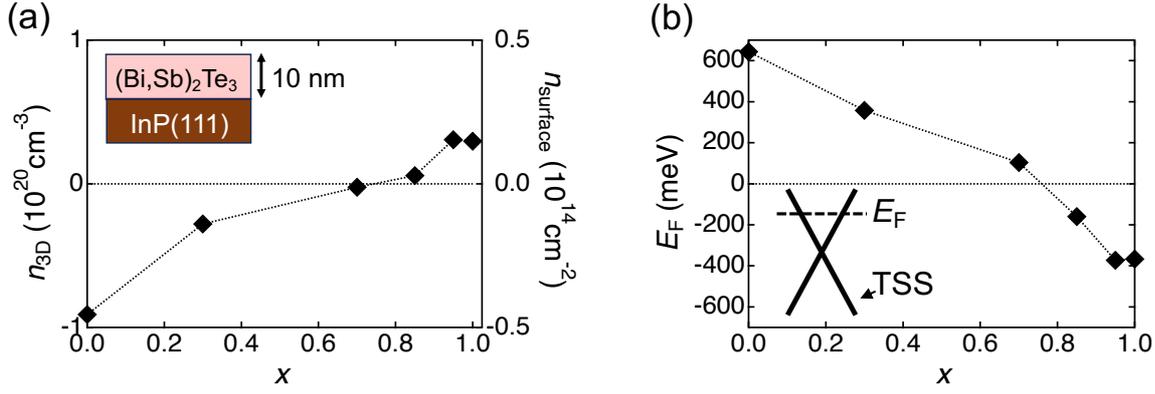

**FIG. S7. Estimated carrier density and Fermi energy in (Bi$_{1-x}$Sb$_x$)$_2$Te$_3$ single-layer film.** (a) Sb composition $x$ dependence of the 3D carrier density $n_{3D}$ in (Bi$_x$Sb$_{1-x}$)$_2$Te$_3$ single-layer films. The carrier density for each surface $n_{surface}$ is also shown in the right column, assuming that (Bi$_x$Sb$_{1-x}$)$_2$Te$_3$ is only composed of top and bottom surface. We calculated $n_{surface}$ from the relation $n_{3D} = 2n_{surface}/d$, where 2 accounts for two surfaces of top and bottom, and $d$ means sample thickness of 10 nm. (b) Sb composition $x$ dependence of Fermi energy $E_F$. We calculated $E_F$ from the relation $E_F = \sqrt{4\pi n_{surface}}\hbar v_F$, where $v_F$ means Fermi velocity. We obtained the value of $v_F = 4.1\times10^5$ m/s from Ref. [17]. The carrier density and Fermi energy are calculated using the data in Fig. 2(h) in the main text. It should be noted that this estimation is valid only around the charge neutral point (CNP) where $E_F = 0$, since Fermi energy crosses bulk bands if it is too far from CNP.

# Supplementary Discussion 1 — AC and DC measurements for the nonreciprocal charge transport

Schematics of AC and DC measurements are illustrated in Fig. S3(a). We apply $I_{AC} = \sqrt{2}I_0 \sin \omega t$ in the AC measurement and $I_{DC} = I_0$ in the DC measurement. According to the formula of nonreciprocal resistance $R = R_0[1 + \gamma \hat{z} \cdot (I/w \times B)]$ (Eq. (1) in the main text), the imaginary part of second harmonic resistance $R_{2\omega}$ in the AC measurement signals NCT which is written as $R_{2\omega} = -R_0 \gamma B I_0 / \sqrt{2}w$. In the DC measurement, the difference in resistance under opposite current directions $\Delta R_{DC}$ signals NCT which is written as $\Delta R_{DC} = 2R_0 \gamma B I/w$. Therefore, $-\sqrt{2}R_{2\omega}$ and $\Delta R_{DC}/2$ should show equal values. We measured NCT of a BT/FST sample by both AC (Fig. S3(b)) and DC (Fig. S3(c)) currents and plotted the magnetic field dependence of $-\sqrt{2}R_{2\omega}$ and $\Delta R_{DC}/2$ in Fig. S3(d). It shows a good agreement especially around zero magnetic field (-0.5 T < $B$ < 0.5 T). The deviation in the high field region may be due to the higher order term in the nonreciprocal resistance which can be captured by the DC measurement. The consistency between the AC and DC measurements assures that the measured $R_{2\omega}$ in the AC measurement is primarily due to the intrinsic NCT.

# Supplementary Discussion 2 — Electron transfer from (Fe,Se)Te to (Bi,Sb)₂Te₃ in (Bi,Sb)₂Te₃/FST heterostructures

The work functions of BT, ST, and FST are reported to be 5.25 eV [28], 5.0 eV [29] and 4.1-4.4 eV [30], respectively. Therefore, we anticipate that electron transfers from FST to BST at the interface of BST/FST. The effect of the electron transfer can be seen in the transport measurement of the heterostructure films. Figures S5(b) and S5(c) show transport properties of single-layer thin films of BT, ST, and FST which are components of the heterostructure films studied in the present paper. Figures S3(a)-S3(d) show transport properties of the heterostructure films BT/FST (solid purple line) and ST/FST (solid orange line). We calculated the resistance and magnetic field derivative of Hall conductance (d$\sigma^{2D}_{xy}$/d$B$) using the data on single-layer films shown in Fig. S5 and plotted them by dotted lines in Figs. S6(a)-(d). Specifically, the dotted line in Fig. S6(a) is calculated from resistance of Bi₂Te₃ (10 nm) ($R_{BT}$) and FST (8 nm) ($R_{FST}$). It is calculated assuming a parallel circuit model $R_{BT}R_{FST}/(R_{BT} + R_{FST})$. The magnetic field derivative of Hall conductance in Fig. S6(c) is calculated as d$\sigma^{2D}_{xy}$ /d$B$(BT) + d$\sigma^{2D}_{xy}$ /d$B$(FST) using the values shown in Fig. S5(c). If each layer of the heterostructure does not affect each other, the parallel circuit model should work. However, clear discrepancy is observed between experimental values (solid lines) and calculated values (dotted lines).

The discrepancy can be accounted for by electron transfer from FST to BT or ST. BT is a *n*-type semiconductor. Therefore, the electron transfer from FST to BT causes an increase in the carrier density in TSS, whereas FST is hardly affected because of its significantly larger carrier density ($\sim 10^{21}$ cm$^{-3}$) than that of BT ($\sim 10^{19}$ cm$^{-3}$). This causes the reduction of the resistance compared to the simple parallel circuit connection of BT and FST. On the other hand, ST is a *p*-type semiconductor. Therefore, the electron transfer in turn causes a decrease in the carrier density of TSS, leading to an increase in resistance. The change in carrier density may be also reflected in the values of $d\sigma^{2D}_{xy}/dB$. According to the simple Drude model, the value of $d\sigma^{2D}_{xy}/dB$ is a sum of $ne\mu^2$ on each band. Assuming that $ne\mu^2$ for the FST layer is not changed by the electron transfer, the increase (decrease) in the carrier density in TSS is expected to cause the increase (decrease) in the absolute value of $d\sigma^{2D}_{xy}/dB$. Thus, the absolute value of $d\sigma^{2D}_{xy}/dB$ is larger (smaller) than the calculated value in BT/FST (ST/FST).